\documentclass[manuscript=article]{achemso}

\usepackage{graphicx}
\usepackage{amsmath}
\usepackage{bm}
\usepackage{textcomp}
\usepackage[english]{babel}
\usepackage[utf8]{inputenc}
\usepackage{times}
\usepackage{cancel}
\usepackage[T1]{fontenc}
\usepackage{color}
\usepackage{here}
\usepackage{amsfonts}
\usepackage[percent]{overpic}
\usepackage{chemformula}
\usepackage[version=4]{mhchem}
\usepackage{hyperref}
\usepackage{url}
\usepackage{mathpazo}
\usepackage{soul}

\SectionNumbersOn

\title{A simple micro-swimmer model 
inspired by the general equation for nonequilibrium
reversible-irreversible coupling}

\author{Andr\'{e}s~C\'{o}rdoba}
\affiliation{Department of Chemical Engineering, 
Universidad de Concepci{\'o}n, Concepci{\'o}n 4030000, Chile}
\email{andcorduri@gmail.com}
\author{Jay~D.~Schieber}
\affiliation{Department of Chemical and Biological Engineering, Department of Physics,
Department of Applied Mathematics, and
Center for Molecular Study of Condensed Soft Matter, Illinois Institute of Technology, 
Chicago, IL 60616, USA}
\author{Tsutomu~Indei}
\affiliation{Global Station for Soft Matter, GI-CoRE, Hokkaido University, 
Sapporo, Hokkaido 060-0808, Japan}

\begin{document}

\begin{abstract}
A simple mean-field micro-swimmer model is presented. The model is inspired by the
nonequilibrium thermodynamics of multi-component fluids that undergo chemical
reactions. These thermodynamics can be rigorously described
in the context of the GENERIC (general equation for the nonequilibrium 
reversible-irreversible coupling) framework. 
More specifically this approach was recently applied to non-ideal polymer solutions
(Indei and Schieber, J. Chem. Phys, 146, 184902, 2017).
One of the species of the solution is an unreactive polymer chain 
represented by the bead-spring model.
Using this detailed description as inspiration we then make several
simplifying assumptions to obtain a mean-field model
for a Janus micro-swimmer. The swimmer model considered here 
consists of a polymer dumbbell in a sea of reactants. One of the beads of the dumbbell 
is allowed to act as a catalyst for a chemical reaction between the reactants.
We show that the mean-squared displacement, MSD, of 
the center of mass of this Janus dumbbell exhibits ballistic behavior at
time scales at which the concentration of reactant is large. 
The time scales at which the ballistic behavior is observed in the MSD
coincides with the time scales at which the cross-correlation
between the swimmer's orientation and the direction of its displacement exhibit a maximum.
Since the swimmer model was inspired by the GENERIC framework it
is possible to ensure that the entropy generation is always positive and therefore
the second law of thermodynamics is obeyed.
\end{abstract}

\noindent This article may be downloaded for personal use only. 
Any other use requires prior permission of the author and AIP Publishing. 
This article appeared in A. C\'{o}rdoba, J. D. Schieber and T. Indei,
J. Chem. Phys. 152, 194902 (2020)
and may be found at: \\ \url{https://doi.org/10.1063/5.0003430}.

%\maketitle

\section{Introduction}

Swimming microorganisms, which include bacteria, algae, and spermatozoa, 
play a fundamental role in most biological processes. 
These swimmers are a special type of active particle, that continuously 
convert local energy into propulsive forces, thereby allowing them to 
move through their surrounding fluid medium. While the size, shape, and
propulsion mechanism vary from one organism to the next, 
they all swim at small Reynolds numbers  \cite{OyamaJPSJ2017, gomez2017tuning, 
LugliJPCC2012}. Microorganisms are able to overcome the thermal randomness of 
their surroundings by harvesting energy to navigate in viscous fluid environments. 
In a similar manner, synthetic colloidal microswimmers are capable of mimicking complex 
biolocomotion by means of simple self-propulsion mechanisms.
Experimentally the speed of active particles can be controlled by {\it e.g.} self-generated 
chemical and thermal gradients, but an in-situ control of the swimming direction remains a
challenge \cite{gomez2017tuning}. 
Artificial self-propelled micro- and nanoengines, or swimmers, have been
increasingly attracting the interest of experimental and theoretical researchers.
Synthetic microswimmers also work at low
Reynolds number, where inertia does not sustain motion 
once the driving force stops, and often
require complicated chemistry in both structural and dynamical terms \cite{LugliJPCC2012}. 

Recent advances in microscopy techniques 
have allowed synthetic microswimmers to be studied experimentally
\cite{EbbensLangmuir2011, gomez2017tuning, Michelin2015, Wilson2013}. 
Janus particle swimmers made by
coating fluorescent polymer beads with hemispheres of platinum 
have been fully characterized using video microscopy to
reveal that they undergo propulsion in hydrogen peroxide fuel
away from the catalytic platinum patch \cite{EbbensLangmuir2011}.
The platinum coating shadows the fluorescence signal from half of each swimmer to
allow the orientation to be observed directly and correlated
quantitatively with the resulting swimming direction  \cite{EbbensLangmuir2011}.
Studies of self-propulsion of half-coated spherical colloids in critical binary
mixtures have shown that the coupling of local body forces, 
induced by laser illumination, and the
wetting properties of the colloid, can be used to finely 
tune both the colloid's swimming speed and its directionality \cite{gomez2017tuning}. 
Among the few methods which have been proposed to create small-scale swimmers, 
those relying on self-phoretic mechanisms present an interesting design 
challenge in that chemical gradients are required to generate net propulsion \cite{Michelin2015}. 
Asymmetries in geometry are sufficient
to induce chemical gradients and swimming  \cite{Michelin2015}. 
Geometric asymmetries can be tuned to induce large
locomotion speeds without the need of chemical patterning \cite{Michelin2015}.
The effect of the fuel concentration on the movement of self-assembled 
nanomotors based on polymersomes has also been studied.
Positive control over the speed of these nanomotors and insights into the
mechanism of propulsion have been reported \cite{Wilson2013}.

Microswimmers have also been studied theoretically and through computer simulations 
\cite{Datt2018, Yasuda2017, Nasouri2017, Milster2017, 
childressJFM2012,sabass2012dynamics, FalascoPRE2016, 
golestanian2005propulsion,
popescu2010phoretic,thakur2011dynamics}.
For instance, catalytic Janus particles
which catalyze a chemical reaction inside the fluid and display 
self-propulsion have been studied extensively using simulations. 
It has been shown that the catalytic reaction produces an asymmetric, 
non-equilibrium distribution of reaction products
around the colloid, which generates osmotic or other
phoretic forces \cite{golestanian2005propulsion, popescu2010phoretic, 
sabass2012dynamics,thakur2011dynamics, LugliJPCC2012}. 
Analogous to the behavior of macroscopic motors the particle 
shape greatly matters and influences not only the
random diffusion but also the efficiency of the propulsive motion \cite{LugliJPCC2012}.
Coarser models of chemically propelled Janus swimmers have also
been previously proposed \cite{Milster2017,OyamaJPSJ2017}.
For example, the dynamical description of the 
motion of a stochastic microswimmer with constant speed and fluctuating
orientation in the long time limit has been 
obtained by adiabatic elimination of the orientational variable \cite{Milster2017}. 
Starting with the corresponding full set of Langevin
equations, the memory in the stochastic orientation is eliminated to
obtain a stochastic equation for the position alone in the overdamped
limit. An equivalent procedure based on the Fokker-Planck equation has
been presented as well \cite{Milster2017}.
Other simulations have used the squirmer
model, which provides an ideal representation of swimmers as spheroidal particles 
that propel owing to a modified boundary condition at their surface.
With this type of model the single-particle and many-particle dynamics of swimmers in
bulk and confined systems have been studied using the smoothed profile method, 
which allows to efficiently solve the coupled particle-fluid problem  \cite{OyamaJPSJ2017}.
Another line of work focuses on the collective behavior that 
emerges in moderately concentrated
suspensions of swimmers \cite{krishnamurthy2015collective,
saintillan2008instabilities, hernandez2009dynamics, simha2002hydrodynamic}.
In those works individual swimmers  are usually modeled using force
dipoles. The hydrodynamic interactions
between swimmers and/or confining walls are treated 
using Stokesian dynamics \cite{brady1988stokesian} 
or similar methods \cite{hernandez2007fast}.

Moreover other theoretical studies have considered different types of swimming mechanisms
\cite{Choudhary2018, OyamaJPSJ2017, LugliJPCC2012,Datt2018}.
For example, swimmers comprising of two rigid spheres 
of different size which oscillate periodically 
along their axis of symmetry, have been shown to propel 
forward in viscoelastic fluids \cite{Datt2018}. 
The dynamics of a generalized three-sphere microswimmer in 
which the spheres are connected by two
elastic springs have also been studied previously. 
The rest length of each spring was assumed to undergo a prescribed 
cyclic change \cite{Yasuda2017}. 
In the low-frequency region,  the swimming velocity increases with frequency. 
Conversely, in the high-frequency region, the average velocity decreases
with increasing frequency \cite{Yasuda2017}. Such behavior originates 
from the intrinsic spring relaxation dynamics of an elastic swimmer
moving in a viscous fluid \cite{Yasuda2017}.
A model swimmer consisting of two linked spheres, 
wherein one sphere is rigid and the other an
incompressible neo-Hookean solid was also proposed \cite{Nasouri2017}. 
The two spheres are connected by a rod which changes its length
periodically. Deformations of the body 
are non-reciprocal despite the reversible
actuation and hence, the elastic two-sphere swimmer propels forward. 
These results indicate that even weak elastic deformations of a body 
can affect locomotion and may be exploited in designing
artificial microswimmers \cite{Nasouri2017}. 
More recent work uses a model one-hinge artificial swimmer consisting of a
bead spring model for two arms joined by a hinge with bending
potential for the arms in order to make them semi-flexible \cite{Choudhary2018}. 
These type of simulations have shown 
that when the swimmer has rigid arms, the center of mass of the swimmer
is not able to propel itself. The flexibility in the
arms causes the time reversal symmetry to break in the case of the one-hinged swimmer 
without the presence of a head \cite{Choudhary2018}. 
The velocity of this type of swimmer has been studied using a range of parameters like
flexibility, beating frequency and the amplitude of the beat \cite{Choudhary2018}.

Thermodynamic analyses of microswimmers have also been attempted
\cite{childressJFM2012,sabass2012dynamics, FalascoPRE2016}. 
In one particular approach a potential thermodynamic 
efficiency was proposed by partially tethering the swimmer so that work is
done externally and instantaneously \cite{childressJFM2012}. 
This instantaneous definition has also been 
extended to encompass a full swimming stroke, and compute it for propulsion of a 
spherical body by a helical flagellum \cite{childressJFM2012}.
Scalar and vectorial steady-state fluctuation theorems and a thermodynamic
uncertainty relation that link the fluctuating particle current 
to its entropy production at an effective temperature have
been proposed \cite{FalascoPRE2016}. Also analytical formulae from linear response theory
have been extended to describe small swimmers, which interact with their environment on a 
finite length scale \cite{sabass2012dynamics}. 
Using irreversible, linear thermodynamics to formulate an 
energy balance the efficiency of transport for swimmers which are moving in random directions
was shown to scale as the inverse of the macroscopic distance over
which transport occurs \cite{sabass2012dynamics}. 

Microswimmers operate far from equilibrium, therefore a complete
and rigorous thermodynamic analysis of systems with microswimmers requires
a thermodynamics framework capable of handling systems far from equilibrium.
GENERIC is an equation to describe the time-evolution
of the state variables not only near equilibrium but far from the
equilibrium state \cite{ottinger2005beyond, 
grmela1997dynamics, ottinger1997dynamics}. 
GENERIC comprises reversible and irreversible 
contributions to the entire dynamics of the system. The
reversible or mechanical contribution is driven by the gradient
of the system energy with respect to the state variables, whereas
the irreversible or dissipative dynamics is driven by the thermodynamic 
forces represented by the gradient of the entropy.
Thus the two generators, the total energy and the total entropy
of the system, drive the full dynamics of the system 
represented by the state variables. Another aspect of GENERIC
is that it can be a useful tool to evaluate the thermodynamic
consistency of coarse-grained models of complex soft materials 
\cite{schieber2003generic, vazquez2009smoothed, wagner1999generalized}.

In  the two-generator formalism of non-equilibrium thermodynamics 
\cite{ottinger2005beyond, grmela1997dynamics, ottinger1997dynamics} 
the time-evolution of a closed system is described by the equation,
\begin{align}\label{dynGENERIC}
\frac{d\bm{x}}{dt}=\bm{L}\cdot\frac{\delta E}{\delta \bm{x}}+
\bm{M}\cdot \frac{\delta S}{\delta \bm{x}}.
\end{align}
Where $\bm{x}$ is a variable that represents the state of the system,
$E$ is the total energy, and $S$ is the total entropy of the system.
eq. (\ref{dynGENERIC}) is called the GENERIC.
The first term describes reversible dynamics. The reversible
dynamics are driven by the gradient of the energy with respect to
the state variable. The operator $\bm{L}$ is called the ``Poisson matrix''.
On the other hand, the second term represents the irreversible
dynamics of the system. The irreversible dynamics are driven
by the entropy gradient with respect to $\bm{x}$. The operator $\bm{M}$ is
referred to as the ``friction matrix''. There are several restrictions 
imposed on $\bm{L}$ and $\bm{M}$. 
First of all, these operators must satisfy the degeneracy conditions:
\begin{align}\label{condGENERIC}
\bm{L}\cdot\frac{\delta S}{ \delta \bm{x}}=\bm{0}, ~~~~~ 
\bm{M}\cdot\frac{\delta E}{ \delta \bm{x}}=\bm{0}.
\end{align}

Intuitively these requirements mean that the entropy gradient
does not generate reversible dynamics whereas the energy gradient 
does not cause irreversible dynamics. In addition, $\bm{L}$ must
be antisymmetric and satisfy the Jacobi identity, whereas $\bm{M}$
must be symmetric and positive semi-definite to guarantee a
positive entropy generation rate or the second law of thermodynamics. 
The entropy generation rate density $\sigma$ can be obtained
by applying the entropy gradient from the left of eq. (\ref{dynGENERIC}) 
and by using the degeneracy requirement for the Poisson matrix,
eq. (\ref{condGENERIC}), as
\begin{align}\label{entropyGENERIC}
\sigma=\frac{\delta S}{\delta \bm{x}}\cdot \bm{M} \cdot \frac{\delta S}{\delta \bm{x}}.
\end{align}

In a recent paper Indei and Schieber \cite{IndeiGENERIC2017}
reexamined the nonequilibrium thermodynamics of multi-component 
fluids that undergo chemical
reactions and showed how to describe it in the context of 
the GENERIC framework. 
They applied this approach to polymer solutions.
One of the species of the solution is the unreactive polymer chain 
represented by the bead-spring model. 
The polymer solution is neither dilute nor ideal. 
The solution entropy was constructed so that the
contributions from mixing and chain conformation are fully separated.

GENERIC is an appropriate tool to check the thermodynamic consistency of a 
set of dynamical equations, even
for active matter. The particular active system that is addressed in this 
paper are self-propelled microswimmers suspended in a Newtonian fluid
\cite{EbbensLangmuir2011, gomez2017tuning}.
For example, Janus polystyrene spheres with a hemisphere 
covered with Pt. A catalytic reaction \ce{H_2O_2 -> 2H_2O + O_2}
occurs at the Pt side of the particle and is the power source
of the propulsion \cite{EbbensLangmuir2011}. Thus \ce{H_2O_2} molecules 
play the role of ``fuel''. This reaction drives the swimmer along a particular direction
inherent to the swimmer’s orientation. 
Though this active system is synthetic, 
this can be a good starting point for understanding more 
complex biological swimmers.
\textcolor{black}{Previous theoretical studies of diffusiophoretic
microswimmers have used more detailed descriptions 
\cite{golestanian2005propulsion, popescu2010phoretic, 
sabass2012dynamics,thakur2011dynamics, LugliJPCC2012}. 
However, a coarse grained minimal level of description
for diffusiophoretic microswimmers still remains to be identified. 
In this paper such a minimal level of description is proposed using
GENERIC as a guide.}

The remainder of this paper is organized as follows: 
Section \ref{model} starts with a brief summary of  the more general
mass balance equations for a non-ideal solution of 
monodisperse polymer chains  dissolved 
in a solvent that is undergoing chemical reactions \cite{IndeiGENERIC2017}.
The assumptions made to obtain the mean-field swimmer model
considered here and the resulting mass balance equations are also given
Section \ref{model}. To perform numerical simulations with 
the Janus swimmer model the mass balance equations are 
written as stochastic differential equations in the 
phase space, these are given at the end of Section \ref{model}.
In Section \ref{entropy} the entropy balance for the 
Janus swimmer model is derived. This balance allows the thermodynamic
compliance of the Janus swimmer model to be readily checked.
Finally, in Section \ref{results}, the predictions of the  
Janus swimmer model are presented and discussed.

\section{Swimmer model} \label{model}

The swimmer model considered here is a bead-spring dumbbell with 
the position of bead $1$ given by $\bm{r}_1$ and the position 
of bead $2$ given by $\bm{r}_2$. The bond vector connecting the two beads is
$\bm{Q}:=\bm{r}_{2}-\bm{r}_{1}$, and the 
center of gravity of the dumbbell is $\bm{r}_c:=\dfrac{\bm{r}_1+\bm{r}_2}{2}$.
The inertia of the beads is neglected and over-damped dynamics are considered
for the equations of motion of the dumbbell. The dumbbell is dissolved in a 
solvent formed by species $\alpha$ and $\beta$. Bead $1$ of the dumbbell
acts as a catalyst for the irreversible chemical reaction $|\nu_\alpha|\alpha  
\longrightarrow |\nu_\beta|\beta$, where 
$\nu_\alpha$ and $\nu_\beta$ are stoichiometric coefficients.

\begin{figure}[h t]
\includegraphics[width=0.4\linewidth]{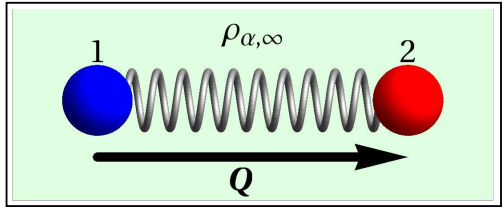}
\caption{Sketch of the simple swimmer model considered in this work.
The swimmer model considered here is a bead-spring dumbbell with 
the position of bead $1$ given by $\bm{r}_1$ and the position 
of bead $2$ given by $\bm{r}_2$, the following definitions
are also used throughout the text,
$\bm{Q}:=\bm{r}_{2}-\bm{r}_{1}$ and $\bm{r}_c:=\dfrac{\bm{r}_1+\bm{r}_2}{2}$.
The radius of both beads is $R$. Bead $1$
acts as a catalyst for the irreversible chemical reaction $|\nu_\alpha|\alpha  
\longrightarrow |\nu_\beta|\beta$.
The swimmer is in a sea (or bath) of reactants and
the concentration of reactant, is equal to $\rho_{\alpha,\infty}$.}
\label{fig1}
\end{figure}

Before giving the simplified equations for the mean-field swimmer model that will
be considered here, the more general mass balance equations for a non-ideal solution of 
monodisperse polymer chains \cite{IndeiGENERIC2017} is briefly recapitulated.
In general, an adequate set of state variables to describe the system discussed above is:
\begin{equation}
\bm{x}(\bm{r},t)=\left\{ \rho_\alpha(\bm{r}), \rho_\beta(\bm{r}), \rho_c(\bm{r},\bm{Q}), 
\bm{u}(\bm{r}), \epsilon(\bm{r})\right\}
\end{equation}
where $\bm{r}$ is a position in space, 
$\rho_\alpha(\bm{r})$ and $\rho_\beta(\bm{r})$ are the mass densities of the 
$\alpha$ and $\beta$ species, respectively; $\rho_c(\bm{r},\bm{Q})$ is the mass density 
of dumbbells; $\bm{u}=\bm{v}\left( \rho_c+\rho_\alpha+\rho_\beta \right)$ 
is the momentum density with 
$\bm{v}$ the velocity; and $\epsilon(\bm{r})$ is the internal energy density.
Note that $\rho_c(\bm{r},\bm{Q})$ may also be interpreted as the probability
density of finding a dumbbell with its 
center of mass at position $\bm{r}$ and with end-to-end vector
$\bm{Q}$. Is this latter interpretation that will be used in the remainder 
of this article.

The mass balance equation for the bead-spring Janus dumbbells is,
\begin{align}\label{balanceS}
\frac{\partial \rho_c(\bm{r},\bm{Q})}{\partial t}=
&-\frac{\partial}{\partial \bm{r}}\cdot
\left[\rho_c(\bm{r},\bm{Q}) \bm{v}(\bm{r})+
\bm{j}_c^{\rm diff}(\bm{r},\bm{Q})
\right] \\ \nonumber
&-\frac{\partial}{\partial \bm{Q}}\cdot \left[ 
\rho_c(\bm{r},\bm{Q}) \nabla\bm{v}(\bm{r})^\dagger\cdot\bm{Q}+
\bm{j}^{\rm conf}(\bm{r},\bm{Q})
\right].
\end{align}
Where $\rho_c(\bm{r},\bm{Q})$ is the mass density of dumbbells. 
The diffusional flux of the dumbbells is given by,
\begin{align}\label{fluxcm}
\bm{j}^{\rm diff}_c(\bm{r},\bm{Q})=&-D'_{c\alpha}(\bm{r},\bm{Q})
\frac{\partial}{\partial \bm{r}}\frac{\mu_\alpha(\bm{r})}{T(\bm{r})}
-D'_{c\beta}(\bm{r},\bm{Q})
\frac{\partial}{\partial \bm{r}}\frac{\mu_\beta(\bm{r})}{T(\bm{r})} \\ \nonumber
&-D'_{cc}(\bm{r},\bm{Q})
\frac{\partial}{\partial \bm{r}}\frac{\mu_c(\bm{r},\bm{Q})}{T(\bm{r})}.
\end{align}
Where $\mu_\alpha(\bm{r})$ and$ \mu_\beta(\bm{r})$ are the chemical
potentials of species $\alpha$ and $\beta$ respectively. 
$D'_{c\alpha}$, $D'_{c\beta}$ and $D'_{cc}$ are diffusion coefficients
and $T$ is the temperature.
The configurational flux of the Janus dumbbells is given by,
\begin{align}\label{fluxconf}
\bm{j}^{\rm conf}(\bm{r},\bm{Q})=&
-4D'_{cc}(\bm{r},\bm{Q})
\frac{\partial}{\partial \bm{Q}}
\left[
\frac{\mu_c(\bm{r},\bm{Q})}{T(\bm{r})}
+\frac{\phi^{(E)}(\bm{Q})}{m_cT(\bm{r})}\right].
\end{align}
Where $\phi^{(E)}(\bm{Q})$ is the 
enthalpic part of the free energy of the spring 
that connects the two beads and $\mu_c(\bm{r},\bm{Q})$ is the chemical potential of a
dumbbell per mass,
\begin{equation}\label{mucrq}
m_c \mu_c(\bm{r},\bm{Q}) = m_c \mu_c(\bm{r})+\phi^{(S)}(\bm{r},\bm{Q})+
k_{\rm B}T(\bm{r})\log \frac{\rho_c(\bm{r},\bm{Q})}{z\rho_c(\bm{r})}.
\end{equation}
Here $m_c$ is the mass of one dumbbell,
$\phi^{S}(\bm{r},\bm{Q})$ is the entropic part of the free energy of the 
spring connecting the two beads, $\rho_c(\bm{r}):=\int \rho_c(\bm{r},\bm{Q})d\bm{Q}$,
and $z$ is a normalization constant.
Note that eqs. (\ref{fluxcm}) and (\ref{fluxconf}) are identical to the
balance equations for a multi-component non-ideal polymer solution
given in previous work \cite{IndeiGENERIC2017}.

The mass balance equations for species $\alpha$ ad $\beta$ are,
\begin{subequations}\label{balanceSp}
\begin{align}
\frac{\partial \rho_\alpha(\bm{r},\bm{Q})}{\partial t}=-\frac{\partial}{\partial \bm{r}}\cdot
\left[\rho_\alpha(\bm{r}) \bm{v}(\bm{r})+
\bm{j}_\alpha^{\rm diff}(\bm{r},\bm{Q})
\right]
+m_\alpha \nu_\alpha J(\bm{r}),
\end{align}
\begin{align}
\frac{\partial \rho_\beta(\bm{r},\bm{Q})}{\partial t}=-\frac{\partial}{\partial \bm{r}}\cdot
\left[\rho_\beta(\bm{r}) \bm{v}(\bm{r})+
\bm{j}_\beta^{\rm diff}(\bm{r},\bm{Q})
\right]
+m_\beta \nu_\beta J(\bm{r}).
\end{align}
\end{subequations}
Where,
\begin{subequations}
\begin{align}
\bm{j}_\alpha^{\rm diff}=&-D'_{\alpha\alpha}(\bm{r})\frac{\partial}{\partial \bm{r}}
\frac{\mu_\alpha(\bm{r})}{T(\bm{r})}
-D'_{\alpha\beta}(\bm{r})\frac{\partial}{\partial \bm{r}}
\frac{\mu_\beta(\bm{r})}{T(\bm{r})}
-D'_{\alpha c}(\bm{r},\bm{Q})\frac{\partial}{\partial \bm{r}}
\frac{\mu_c(\bm{r})}{T(\bm{r})},
\end{align}
\begin{align}
\bm{j}_\beta^{\rm diff}=&-D'_{\beta\beta}(\bm{r})\frac{\partial}{\partial \bm{r}}
\frac{\mu_\beta(\bm{r})}{T(\bm{r})}
-D'_{\beta\alpha}(\bm{r})\frac{\partial}{\partial \bm{r}}
\frac{\mu_\alpha(\bm{r})}{T(\bm{r})}
-D'_{\beta c}(\bm{r},\bm{Q})\frac{\partial}{\partial \bm{r}}
\frac{\mu_c(\bm{r})}{T(\bm{r})}.
\end{align}
\end{subequations}
$J(\bm{r})$ is the reaction rate and $\mu_c(\bm{r}):=\int \mu_c(\bm{r},\bm{Q})d\bm{Q}$. 
Here also the mass balance equations are the same as the ones 
previously derived for a multi-component 
non-ideal polymer solution \cite{IndeiGENERIC2017}. Moreover,
since the total mass flux has to be conserved the diffusion coefficients 
$D'_{c\alpha}$, $D'_{c\beta}$
are coupled to $D'_{cc}$ through \cite{IndeiGENERIC2017},
\begin{equation}
D'_{cc}(\bm{r},\bm{Q})+D'_{c\alpha}(\bm{r},\bm{Q})
+D'_{c\beta}(\bm{r},\bm{Q})=0.
\end{equation}

\textcolor{black}{
Additionally to the mass balance equations for the solvent species, eq. (\ref{balanceSp}),
and the mass balance for the dumbbell, eq. (\ref{balanceS}), balance equations for the 
remaining state variables, the
momentum density and the total energy density, have also been previously obtained
using GENERIC for the non-ideal solution of monodisperse polymer chains \cite{IndeiGENERIC2017}.
However they will not be used in this manuscript 
and therefore they are not rewritten here. }

Note that in general  eqs. (\ref{balanceSp}) and (\ref{balanceS}) represent a
system of coupled equations for $\rho_\alpha(\bm{r})$, $\rho_\beta(\bm{r})$ and
$\rho_c(\bm{r},\bm{Q})$. In principle, after specifying 
a reaction rate $J(\bm{r})$ this system of equations together with the balances for
$\bm{u}(\bm{r})$ and $\epsilon(\bm{r})$
can be solved numerically for some
given boundary conditions. In this work a different approach is taken, 
here the purpose is to obtain a mean-field model of a
single Janus microswimmer in which the thermodynamic
compliance can be readily checked.
The swimmer is in a sea (or bath) of reactants and
the concentration of reactant in the bath is equal to $\rho_{\alpha,\infty}$.
Note that $\rho_{\alpha,\infty}$ will be allowed to change with 
time, but it will not be a field variable.
Fig. \ref{fig1} shows a sketch of the Janus swimmer model considered here.
The level of description (set of state variables) of the model then reduces to,
\begin{equation}\label{lofds}
\bm{x}(\bm{r}_c,\bm{Q})=\left\{ \rho_{\alpha,\infty}, \rho_c(\bm{r}_c,\bm{Q}), 
\epsilon(\bm{r}_c)\right\}.
\end{equation}

Moreover it will be assumed that very near bead $1$ of the dumbbell, 
{\it i.e.} the catalytic bead, the concentration of species $\alpha$ 
is always very small and will be approximated to be always zero. 
At a distance $\ell_0$ away, near bead $2$ of the dumbbell,
the concentration of of species $\alpha$ will be assumed
to always be equal to $\rho_{\alpha,\infty}$.
Therefore the concentration gradient for $\alpha$ will be in the direction 
of $\bm{Q}$ and can be approximated as,
\begin{align}\label{aproxgrada}
&\dfrac{\partial \rho_\alpha(\bm{r}_c)}{\partial \bm{r}_c}\approx
\dfrac{\bm{Q} }{Q} \frac{\rho_{\alpha,\infty}}{\ell_0} 
\frac{\rho_c(\bm{r}_c,\bm{Q})}{\rho_c}.
\end{align}
Where $Q:=|\bm{Q}|$ is the magnitude of $\bm{Q}$. 
Since a single swimmer in a mean-field is being considered then 
from now on $\dfrac{\rho_c(\bm{r}_c,\bm{Q})}{\rho_c}$ is interpreted as the probability 
density of finding a dumbbell with its center of mass at position $\bm{r}_c$ and
end-to-end vector $\bm{Q}$.
With $\rho_c:= \dfrac{\int \int \rho_c(\bm{r}_c,\bm{Q})d\bm{Q}d\bm{r}_c}{V}=\dfrac{m_c}{V}$
being the normalization factor for the probability density.
And $V$ is the total volume of the container where the dumbbell is swimming.
Similarly, across the swimmer there will also be a concentration 
gradient for species $\beta$ that can be approximated as,
\begin{align}\label{aproxgradb}
\dfrac{\partial \rho_\beta(\bm{r}_c)}{\partial \bm{r}_c}&\approx
-\dfrac{\bm{Q} }{Q} \frac{1}{\ell_0} \left(\rho_{\beta,\infty}+
\frac{|\nu_\beta|}{|\nu_\alpha|}\frac{m_\beta}{m_\alpha}
\rho_{\alpha,\infty}-\rho_{\beta,\infty}\right) \frac{\rho_c(\bm{r}_c,\bm{Q})}{\rho_c} \\ \nonumber
&=-\dfrac{\bm{Q} }{Q} \frac{\rho_{\alpha,\infty}}{\ell_0} 
\frac{\rho_c(\bm{r}_c,\bm{Q})}{\rho_c}.
\end{align}
Where $\rho_{\beta,\infty}$ is the concentration 
of $\beta$ near bead $2$ of the dumbbell, and
$\rho_{\beta,\infty}+\dfrac{|\nu_\beta|}{|\nu_\alpha|}\dfrac{m_\beta}{m_\alpha}\rho_{\alpha,\infty}$ 
is the concentration of $\beta$
near the catalytic bead of the dumbbell, {\it i.e.} bead $1$; 
$m_\alpha$ and $m_\beta$ are the molecular weights of 
$\alpha$ and $\beta$ respectively.  In the second line of eq. (\ref{aproxgradb}) the fact that
$\dfrac{|\nu_\beta|}{|\nu_\alpha|}\dfrac{m_\beta}{m_\alpha}=1$
due to mass conservation has been used.

With the approximations for the concentration
gradients of $\alpha$ and $\beta$ near the swimmer, 
eqs. (\ref{aproxgrada}) and (\ref{aproxgradb}),  assuming constant temperature
and using the chain rule for the gradients of the chemical potentials, {\it e.g.}
$\frac{\partial \mu_\alpha(\bm{r})}{\partial \bm{r}}
=\bigg\{\frac{\partial \rho_\alpha(\bm{r})}{\partial \bm{r}}
\left[\frac{\partial \mu_\alpha(\bm{r})}
{\partial \rho_\alpha(\bm{r})}\right]_{T,p,\rho_\beta,\rho_c}
+\frac{\partial \rho_\beta(\bm{r})}{\partial \bm{r}}
\left[\frac{\partial \mu_\alpha(\bm{r})}
{\partial \rho_\beta(\bm{r})}\right]_{T,p,\rho_\alpha,\rho_c} 
+\frac{\partial \rho_c(\bm{r},\bm{Q})}{\partial \bm{r}}
\left[\frac{\partial \mu_\alpha(\bm{r})}
{\partial \rho_c(\bm{r},\bm{Q})}\right]_{T,p,\rho_\alpha,\rho_\beta}
\bigg\}$ the diffusional 
flux of the Janus dumbbell, eq. (\ref{fluxcm}), can be written as,
\begin{align}\label{difsimp}
\bm{j}^{\rm diff}_c(\bm{r}_c,\bm{Q})=&-
\bigg[D'_{c\alpha}(\bm{r}_c,\bm{Q})\gamma'_\alpha(\bm{r}_c,\bm{Q})+
D'_{c\beta}(\bm{r}_c,\bm{Q})\gamma'_\beta(\bm{r}_c,\bm{Q})\bigg]
\frac{\bm{Q}}{Q}\frac{\rho_{\alpha,\infty}}{\ell_0}\frac{\rho_c(\bm{r}_c,\bm{Q})}{\rho_c} \\
\nonumber
&-\bigg[ D'_{c\alpha}(\bm{r}_c,\bm{Q}) \gamma'_{c\alpha}(\bm{r}_c,\bm{Q})+
D'_{c\beta}(\bm{r}_c,\bm{Q})  \gamma'_{c\beta}(\bm{r}_c,\bm{Q})\bigg]
\frac{\partial \rho_c(\bm{r}_c,\bm{Q})}{\partial \bm{r}_c}.
\end{align}
Where $\gamma'_\alpha(\bm{r}_c,\bm{Q})$, $\gamma'_\beta(\bm{r}_c,\bm{Q})$,
$\gamma'_{c\alpha}(\bm{r}_c,\bm{Q})$ and $\gamma'_{c\beta}(\bm{r}_c,\bm{Q})$
are thermodynamic derivatives of the chemical potentials with respect to the 
densities. The definitions for $\gamma'_\alpha(\bm{r}_c,\bm{Q})$, $\gamma'_\beta(\bm{r}_c,\bm{Q})$,
$\gamma'_{c\alpha}(\bm{r}_c,\bm{Q})$ and $\gamma'_{c\beta}(\bm{r}_c,\bm{Q})$  are given in 
Section S1 of the Supplementary Information.
Furthermore, assuming that the spring connecting the two beads is purely entropic 
the configurational flux of the dumbbell, eq. (\ref{fluxconf}), can be written as,
\begin{align}\label{confsimp}
\bm{j}^{\rm conf}(\bm{r}_c,\bm{Q})=&
-4 \gamma'_{cc}(\bm{r}_c,\bm{Q})\bigg[ D'_{c\alpha}(\bm{r}_c,\bm{Q})+
D'_{c\beta}(\bm{r}_c,\bm{Q})\bigg]
\left[\frac{\partial\rho_c(\bm{r}_c,\bm{Q})}{\partial \bm{Q}} +
\frac{m_c}{k_{\rm B}T}\frac{\partial\phi^{(S)}(\bm{Q})}{\partial \bm{Q}} \right].
\end{align}
Where it has been assumed that the spring connecting the two beads
is purely entropic and therefore $\phi^{(S)}(\bm{Q})$ is all the free energy of the 
spring. Also, in what follows $D'_{c\alpha}$, $D'_{c\beta}$, $\gamma'_{\alpha}$, 
$\gamma'_\beta$, $\gamma'_{c\alpha}$, $\gamma'_{c\beta}$ and 
$\gamma'_{cc}$ will be assumed to be independent of $\bm{r}_c$ and $\bm{Q}$.
Note that since near the swimmer the mass fluxes of
the two solvent species are equal in magnitude but opposite in 
direction. Then, the terms in the mass balance 
equation for the dumbbell, eq. (\ref{balanceS}), involving the 
velocity, $\bm{v}(\bm{r}_c)$, vanish {\it i.e.},
$\rho_c(\bm{r}_c,\bm{Q}) \bm{v}(\bm{r}_c)=0$ and
$\rho_c(\bm{r}_c,\bm{Q}) \nabla \bm{v}(\bm{r}_c)^\dagger=0$.

The approximations for the gradients 
of the reactants near the swimmer,  eq. (\ref{aproxgrada})
and eq. (\ref{aproxgradb}), effectively 
decouple the mass balance equation for the dumbbell from 
the mass balance equations for the reactants. 
However, for the level of description of the mean-field 
swimmer model one can write an evolution equation
for the concentration of reactant in the bath,
\begin{equation}
\frac{d\rho_{\alpha,\infty}(t)}{dt}=-J(t).
\end{equation}
Where the net reaction rate $J(t)$ now includes both 
the mass transport of reactants from the bath
to the dumbbell and also the 
reaction rate once the reactants reach the catalyst-coated 
bead of the Janus dumbbell.  
Therefore the net reaction rate, $J(t)$, can be written as,
\begin{align}\label{rratesimp}
J(t)=\left(\frac{4 \pi R^2}{V}\right)\frac{1}{1/J_r(t)+1/J_m(t)}.
\end{align}
Where $J_r(t)$ is the rate of reaction at the surface of the
catalyst coated bead and $J_m(t)$ 
is the rate associated with the mass transport of reactants 
from the bulk to the swimmer. $R$ is the radius of the 
catalyst-coated bead and $V$ is the volume of the 
container where the Janus dumbbell is swimming.
The specific form for the reaction rate used in eq. (\ref{rratesimp})
is based on the solution to a problem with similar physics. 
That is, binary diffusion through a boundary layer and 
reaction at a surface \cite{bird2007transport}. See Section S3 of the
Supplementary Information for more details on this regard.
In general, for the reaction rate, $J_r(t)$, the following
form can be employed,
\begin{align}
J_r(t)=|\nu_\alpha| k'' \rho_{\alpha,\infty}(t)^{|\nu_\alpha|}.
\end{align}
Where $k''$ is a rate constant for the reaction at the surface of the
catalyst-coated bead of the Janus dumbbell.
Here, we will consider only first order reactions, 
that is $\nu_\alpha=-1$. 
Assuming $D'_{\alpha\alpha}$ and $\gamma_{\alpha \alpha}$ independent 
of $\bm{r}$, and using the approximation for the gradients introduced in 
eqs. (\ref{aproxgrada}) and (\ref{aproxgradb})
the rate associated with mass transport of reactants 
from the bulk to the swimmer may be written as,
\begin{align}
J_m(t)=D'_{\alpha\alpha} \gamma_{\alpha \alpha \beta}
\frac{\rho_{\alpha,\infty}(t)}{\ell_0}.
\end{align}
Therefore the net reaction rate can be written as,
\begin{align}\label{jrate}
J(t)=\left(\frac{4 \pi R^2}{V}\right)
\frac{k'' \gamma_{\alpha \alpha \beta} D'_{\alpha\alpha}
\rho_{\alpha,\infty}(t)}
{\gamma_{\alpha \alpha \beta} D'_{\alpha\alpha}+\ell_0 k''}.
\end{align}
Is also useful to write  the net reaction rate, eq. (\ref{jrate}) in dimensionless form,
\begin{align}
\hat{J}(t)=X_S \left( \frac{X_\rho(t) X_r}{X_r \hat{\ell}_0+1} \right).
\end{align}
Where $X_r:=\dfrac{k'' }{D'_{\alpha\alpha}\gamma_{\alpha\alpha\beta}}
\sqrt{\dfrac{k_{\rm B}T}{k_b}}$
and $X_S:=\dfrac{4 \pi R^2}{V}\sqrt{\dfrac{k_{\rm B}T}{k_b}}$.
Note that $X_r$ is a ratio between two time scales. The time scale
for the mass transfer of reactant from the bath to the dumbbell,
in the numerator, and in the denominator the
time scale of the reaction at the surface of the catalytic bead. 
Therefore values of $X_r>1$ correspond to an overall reaction rate that is more limited by 
the mass transport of reactants
while values of $X_r<1$ correspond to a reaction rate that is more  
limited by the chemical kinetics.

Using the simplified fluxes, eqs. (\ref{difsimp}) and (\ref{confsimp}),
it is now possible to write the mass balance for the swimmer,
eq. (\ref{balanceS}), in the phase space as a set of stochastic
differential equations (SDEs) for $\bm{Q}$ and $\bm{r}_c$ 
\cite{Gardiner2009, Ottinger},
\begin{align}\label{modelf}
d\bm{Q}=&\frac{2\gamma'_{cc}\left(D'_{c\alpha}+D'_{c\beta}\right)}{k_{\rm B} T} 
\frac{\partial \phi^{(S)}}{\partial \bm{Q}}dt +\sqrt{4\gamma'_{cc}\left(D'_{c\alpha}+D'_{c\beta}\right)}
d\bm{W}_1(t),\\ \nonumber
d\bm{r}_c=&\left[\frac{\rho_{\alpha,\infty}(t)}{\ell_0 \rho_c}
\left(D'_{c\alpha}\gamma'_\alpha+
D'_{c\beta}\gamma'_\beta\right)\frac{\bm{Q}}{Q}\right]dt+
\sqrt{ D'_{c\alpha} \gamma'_{c\alpha}+
D'_{c\beta} \gamma'_{c\beta}}d\bm{W}_2(t).
\end{align}
Where $d\bm{W}_1$ and $d\bm{W}_2$ are vectors of Wiener increments
with statistics, 
$\left\langle {d \bm{W}}_{i}(t)\right\rangle_{\rm eq}=\bm{0}$ and 
$\left\langle {d\bm{W}}_{i}(t){d\bm{W}}_{j}(t')\right\rangle_{\rm eq}
=\bm{\delta}\delta(t-t')\delta_{ij}dtdt'$. 
In what follows it will be assumed that $\phi^{(S)}=-\dfrac{k_b (Q-\ell_0)^2}{2}$, 
with $k_b$ the entropic spring constant. Note that in eq. (\ref{modelf}) the first term on the 
right hand side of the equation for the center of mass of the dumbbell 
can be interpreted as the swimming velocity. The swimming velocity
is directly proportional to the concentration of reactant in the bath, $\rho_{\alpha,\infty}$, 
and  inversely proportional to the rest length of the dumbbell, $\ell_0$. Shorter 
dumbbells produce larger concentrations gradients around them, which
in turn produces a larger swimming velocity.
To perform Brownian dynamics (BD) simulations of the swimmer model 
eq. (\ref{modelf}) is first made dimensionless. The dimensionless form 
of eq. (\ref{modelf}) is given in the Supplementary Information, eq. (S17).
In what follows dimensionless variables are written with a hat.

\section{Entropy Balance}\label{entropy}

In the GENERIC framework the entropy generation rate 
density of a model can be calculated using eq. (\ref{entropyGENERIC}).
In general, the entropy generation rate should always be positive
to guarantee that the model satisfies the second law of thermodynamics.
Assuming constant temperature and using the chain rule for the 
chemical potential gradients it can be shown, see Sec. S2 in the Supplementary 
Information, that the entropy generation rate density 
for the dumbbell in a bath of reactants reduces to,
\begin{align}\label{entropygenat}
\sigma = &
\frac{\mu_{\alpha,\infty}(t)}{T} J(t) +
\frac{A_c }{VT}\int\int 
\left[\frac{\rho_{\alpha,\infty}}{\ell_0} 
\frac{\rho_c(\bm{r}_c,\bm{Q})}{\rho_c} \right]^2 d\bm{Q} d\bm{r}_c  \\ \nonumber
+& \frac{C_c }{VT}\int\int \left[
 \dfrac{\partial \rho_c(\bm{r}_c,\bm{Q})}{\partial \bm{r}_c} \cdot
\dfrac{\partial \rho_c(\bm{r}_c,\bm{Q})}{\partial \bm{r}_c}  \right]
d\bm{Q} d\bm{r}_c  \\ \nonumber
+& \frac{4 \gamma^2_{cc} (D'_{c\alpha}+D'_{c\beta})}{VT} \int \int 
\left[\frac{\partial \rho_c(\bm{r}_c,\bm{Q})}{\partial \bm{Q}} 
\cdot \frac{\partial \rho_c(\bm{r}_c,\bm{Q})}{\partial \bm{Q}}\right] d\bm{Q}d\bm{r}_c.
\end{align}
Where $\mu_{\alpha, \infty}$ is
the chemical potential of reactant in the bath. Note that the
first term in eq. (\ref{entropygenat}) includes the entropy generation
due to the diffusion of reactant from the bath to the swimmer and
also includes the entropy generation due to the chemical reaction at the 
surface of the catalytic bead. 
If $\mu_{\alpha, \infty}$ is assumed to depend linearly on $\rho_{\alpha, \infty}$
then this term is always positive for the reaction flux, eq. (\ref{jrate}), used here. 
The term in the last line of eq. (\ref{entropygenat}), is quadratic in 
$\dfrac{\partial \rho_c(\bm{r}_c,\bm{Q})}{\partial \bm{Q}}$
and therefore always positive as long as $D'_{c\alpha}$ and $D'_{c\beta}$ are positive.
Also the following parameters were introduced,
\begin{align}
A_c:=&
D'_{c\alpha} \gamma_{c\alpha\beta} \left[\gamma_{\alpha\alpha\beta}
-\gamma_{c\alpha\beta} \right]+
D'_{c\beta} \gamma_{c\alpha\beta} \left[\gamma_{\beta\alpha\beta}
-\gamma_{c\alpha\beta} \right] \\ \nonumber
C_c:=&D'_{c\alpha}\gamma_{cc}\left[\gamma_{\alpha c} - \gamma_{cc}\right]+ 
D'_{c\beta}\gamma_{cc}\left[\gamma_{\beta c}-\gamma_{cc}\right].
\end{align}
The definitions for the thermodynamic derivatives, 
$\gamma_{c\alpha\beta}$, $\gamma_{\alpha\alpha\beta}$, 
$\gamma_{\beta\alpha\beta}$, $\gamma_{cc}$, $\gamma_{\alpha c}$ and
$\gamma_{\beta c}$,
are given in eq. (S2) of the Supplementary Information.
The remaining terms on the right side of eq. (\ref{entropygenat})
always stay positive as long as $A_c>0$ and $C_c>0$. 
Since $D'_{c\alpha}$ and $D'_{c\beta}$ are positive then the restrictions 
on $A_c$ and $C_c$ translate into restrictions on the $\gamma$'s. 
Is illustrative to assume $\gamma_{\beta\alpha\beta}=\gamma_{\alpha\alpha\beta}$
and $\gamma_{\beta c}=\gamma_{\alpha c}$
in which case those restrictions simplify to,
\begin{align}
\gamma_{\alpha\alpha\beta} > \gamma_{c\alpha\beta}, ~~~~
\gamma_{\alpha c} > \gamma_{c c}.
%, ~~~\frac{\gamma_{\alpha c}}{\gamma_{cc}} + 
%\frac{\gamma_{\alpha\alpha\beta}}{\gamma_{c\alpha\beta}}>1.
\end{align}
Or in terms of $\gamma'_\alpha$, $\gamma'_\beta$, $\gamma'_{c\alpha}$,
and $\gamma'_{c\beta}$ the restrictions are,
\begin{align}\label{conds}
\gamma'_\alpha>0, ~~~ \gamma'_{c\alpha}>0,~~~~
\gamma'_\beta=\gamma'_\alpha, ~~~\gamma'_{c\beta}=\gamma'_{c\alpha}.
\end{align}

In what follows the simplified restrictions given in eq. (\ref{conds}) will be used
to guide the selection of parameters when performing numerical 
simulations of the model.
Note that, in general, the entropy balance for the mean-field Janus dumbbell 
model imposes that the swimming velocity be in the positive direction 
of $\bm{Q}$. Where $\bm{Q}$ points from the catalytic bead towards the 
non-catalytic bead of the dumbbell.
This is in agreement with experimental observations that show
that Janus colloidal particles in hydrogen peroxide 
always swim away from the catalytic platinum 
coated patch \cite{EbbensLangmuir2011}.

Note that here we have used the chemical potential of a passive
dumbbell, eq. (\ref{mucrq}), without modifications and that the
symmetry was broken through the expressions introduced for 
the concentration gradients of reactants 
eqs. (\ref{aproxgrada}) and (\ref{aproxgradb}). It is also
possible to modify the chemical potential of the dumbbell to 
explicitly account for its asymmetry, that is that one of the beads is catalytic and the
other one is not. For example, the following additional term may be added to the
chemical potential of the dumbbell,
$\dfrac{\gamma_c}{2} \bm{Q} \cdot
\dfrac{\partial}{\partial \bm{r}} \dfrac{\partial \rho_{\alpha}(\bm{r})}{\partial \bm{r}}
\approx \dfrac{\gamma_c}{2} \bm{Q} \cdot \dfrac{\rho_{\alpha,\infty}}{\ell_0} \bm{Q}\bm{Q}
=\dfrac{\gamma_c Q^2 \rho_{\alpha,\infty}}{2\ell_0}\bm{Q}$.
Where $\gamma_c$ would be an additional model parameter. Note that the same approximation 
for the gradient of the reactant as used in eq. (\ref{aproxgrada}) was employed
to obtain the final form. This type of additional term in $\mu_c(\bm{r},\bm{Q})$
would result in an additional $\bm{Q}$ dependent term in the configurational
flux of the dumbbell, eq. ($\ref{confsimp}$) and therefore an additional
$\bm{Q}$ dependent term in the SDE for $\bm{Q}$.
The additional term in $\mu_c(\bm{r},\bm{Q})$ would also change the 
prefactor in the swimming velocity.

\section{Results and Discussion}\label{results}

Brownian dynamics (BD) simulations 
are employed to study the dynamics of the Janus swimmer model 
presented in this work. Two quantities are used to monitor the motion
of the Janus dumbbell during the BD simulations. The mean-squared displacement 
of the center of mass,
\begin{align}
\langle \Delta \hat{r}_c^2(t) \rangle=
\langle \left[ \hat{\bm{r}}_c(t)-\hat{\bm{r}}_c(0) \right]^2 \rangle.
\end{align}
Where the brackets, $\langle ...  \rangle$, denote an ensemble average.
Another observable that is useful when tracking the motion
of the Janus dumbbell is the  normalized cross-correlation between the 
direction of $\hat{\bm{Q}}$ and the instantaneous
displacement vector of the swimmer, $\Delta \hat{\bm{r}}(t)$, 
\begin{align}
S(t)=\left\langle \frac{\Delta \hat{\bm{r}}_c(t)}{|\Delta \hat{\bm{r}}_c(t)|}\cdot 
\frac{\hat{\bm{Q}}(0)}{|\hat{\bm{Q}}(0)|}\right\rangle
=\left\langle \frac{\left(\hat{\bm{r}}_c(t)-\hat{\bm{r}_c}(0) \right)}
{|\hat{\bm{r}}_c(t)-\hat{\bm{r}}_c(0)|}
\cdot \frac{\hat{\bm{Q}}(0)}{|\hat{\bm{Q}}(0)|}\right\rangle.
\end{align}
Where $|\Delta \hat{\bm{r}}_c(t)|=\sqrt{\Delta \hat{\bm{r}}_c(t)\cdot\Delta \hat{\bm{r}}_c(t)}$ 
and $|\hat{\bm{Q}}(0)|=\sqrt{\hat{\bm{Q}}(0)\cdot\hat{\bm{Q}}(0)}$. 

\begin{figure}[h t]
\center
\includegraphics[width=\linewidth]{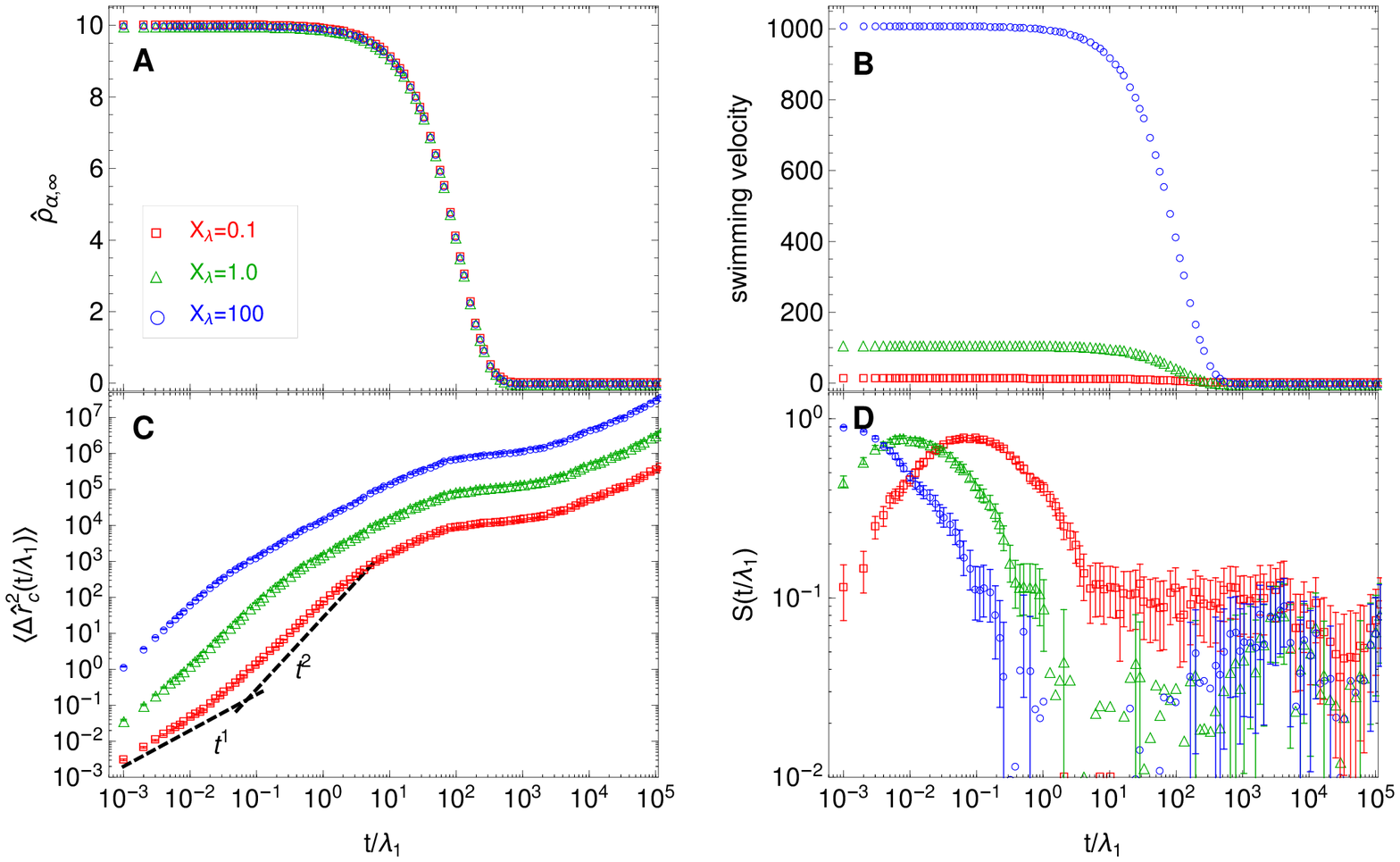}
\caption{Effect of $X_\lambda:=D'_{c\beta}/D'_{c\alpha}$ 
on the swimming dynamics.
(A) Dimensionless concentrations of the reactant in the bath, 
$\hat{\rho}_{\alpha,\infty}$. 
(B) Dimensionless swimming velocity.
(C) Mean-squared displacement of the center of mass of the Janus
swimmer.  (D) Normalized cross-correlation between the 
orientation of the Janus swimmer, $\bm{Q}/Q$, 
and its instantaneous displacement vector, $\Delta \bm{r}(t)$.
Other parameters where set to,
$\hat{\ell}_0=1$, $X_r=10$, $X_S=0.01$.}
\label{fig2}
\end{figure}

Fig. \ref{fig2} shows the effect of the ratio of the diffusion coefficients 
of product and reactant,
$X_\lambda:=D'_{c\beta}/D'_{c\alpha}$ ,
on the dynamics of the dumbbell. Fig. \ref{fig2}A shows the dimensionless 
concentration of the reactant in the bath as a function of time. \textcolor{black}{
As expected for the type of reaction rate used here the concentration
of reactant decreases exponentially with time as the reaction proceeds.}
Fig. \ref{fig2}B shows the swimming velocity,
{\textit i.e.} $\dfrac{\rho_{\alpha,\infty}}{\ell_0 \rho_c}
\left(D'_{c\alpha}\gamma'_\alpha+
D'_{c\beta}\gamma'_\beta\right)$,  as a function
of time for different values of $X_\lambda$. \textcolor{black}{In general, 
since the swimming velocity is proportional to $\rho_{\alpha,\infty}$
it also decreases exponentially towards zero as the reactant is consumed.
But, larger values of $X_\lambda$ produce larger swimming velocities.
Fig. \ref{fig2}C shows the mean-squared displacement of 
the center of mass of the swimmer as a function of time,
$\langle \Delta \hat{r}_c^2(t) \rangle$,
where time has been made dimensionless by 
$\lambda_1:=\dfrac{k_{\rm B}T}{k_bD'_{c\alpha}\gamma'_\alpha}$.}
For values of $X_\lambda$ larger than one, which produce larger 
swimming velocities, $\langle \Delta \hat{r}_c^2(t) \rangle$ 
shows an overall larger magnitude.
In general, for values of $X_\lambda$ smaller or close to one 
$\langle \Delta \hat{r}_c^2(t) \rangle$ exhibits four characteristic regions.
A diffusive region, $\langle \Delta \hat{r}_c^2(t) \rangle \sim t^1$, is observed at short times. 
This is followed by a ballistic region where $\langle \Delta \hat{r}_c^2(t) \rangle\sim t^{2}$.
\textcolor{black}{The ballistic region is followed by 
another diffusive region where the diffusivity of 
the dumbbell is larger than in the first diffusive region by a factor 
that can range from 5 to 17.} Note that this second diffusive region 
appears at times at which the concentration of reactant and 
the swimming velocity are still somewhat large. However due to tumbling,
{\it i.e.} change in the direction of $\bm{Q}$ due to Brownian forces, the
swimmer can not maintain the ballistic motion for a longer amount of time.
\textcolor{black}{Following this second diffusive region a plateau is observed in the 
mean-squared displacement of the center of mass of the swimmer.
This region arises when the concentration of reactant is very small, 
the reactant is near depletion but not quite yet depleted. At this point
the swimming velocity, \textit{i.e.} the term inside the square brackets 
in the second line of eq. (\ref{modelf}), which is proportional to 
the orientation of the dumbbell, $\bm{Q}/Q$, is still finite but becomes very small
and effectively acts like a trap for the center of mass of the dumbbell.
In other words, for small but still finite concentration of reactant 
the center of mass, $\bm{r}_c(t)$, is still coupled to the orientation of 
the dumbbell, $\bm{Q}/Q$. At this point while $\bm{Q}/Q$
keeps changing due to Brownian forces the center of mass, $\bm{r}_c(t)$, moves very little 
in every new direction of $\bm{Q}$, so effectively the dumbbell appears trapped.
When the reactant is completely consumed, \textit{i.e.} $\rho_{\alpha,\infty}=0$, 
the dynamics of the center of mass, $\bm{r}_c(t)$, become completely decoupled from 
the $\bm{Q}(t)$ dynamics and then a final diffusive region of 
purely thermal nature is observed. In this diffusive region that appears 
at long times the diffusivity of the center of mass of the dumbbell is smaller
than the diffusivity observed in the second diffusive region by a factor of about 27.} 
\textcolor{black}{Fig. \ref{fig2}D shows the 
effect of $X_\lambda$ on the cross-correlation, $S(t)$.
It can be observed that in general $S(t)$ exhibits a maximum at 
$t=\lambda_{\rm peak}$
that coincides with the time at which 
$\langle \Delta \hat{r}_c^2(t) \rangle$ shows ballistic behavior.
Therefore, as $X_\lambda$ is increased 
and the ballistic region in $\langle \Delta \hat{r}_c^2(t) \rangle$
moves to longer times so does the maximum in the $S(t)$.
For values of $X_\lambda$ larger than one, the short time
diffusive region in $\langle \Delta \hat{r}_c^2(t) \rangle$ can not be observed, 
this is because the maximum in $S(t)$ moves to very short times and therefore ballistic motion
dominates the $\langle \Delta \hat{r}_c^2(t) \rangle$ at those short times.
For values of $X_\lambda$ smaller than one,
when $D'_{c\alpha}$ is significantly larger than $D'_{c\beta}$, the 
gradient of reactant across the dumbbell tends to be smaller because 
the reactant diffuses more than the product. Therefore  
it takes longer for the reactant gradient across the dumbbell to be large enough to 
sustain swimming motion and therefore the short diffusive region can be
observed at short times.}

\begin{figure}[h t]
\center
\includegraphics[width=\linewidth]{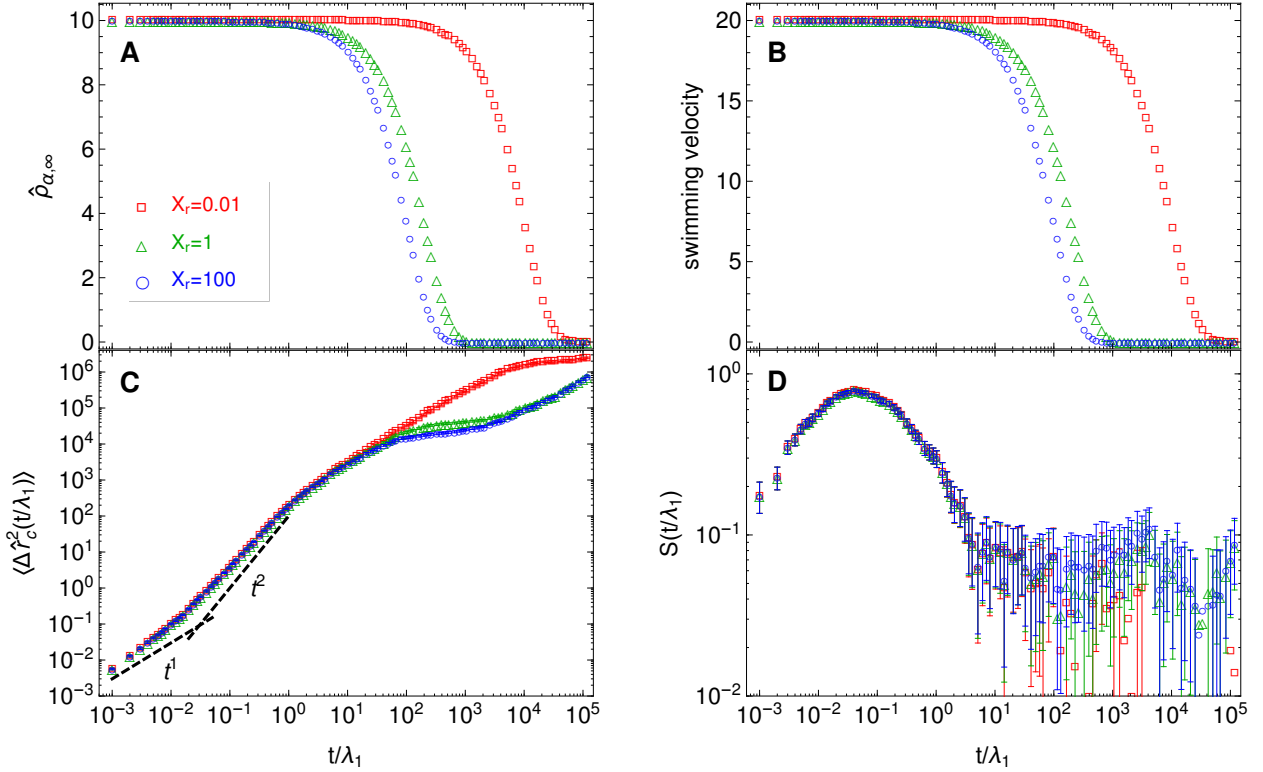}
\caption{Effect of $X_r$ on the swimming dynamics.
(A) Dimensionless concentration of the reactant in the bath, 
$\hat{\rho}_{\alpha,\infty}$. 
(B) Dimensionless swimming velocity.
(C) Mean-squared displacement of the center of mass of the Janus
swimmer.  (D) Normalized cross-correlation between the 
orientation of the Janus swimmer, $\bm{Q}/Q$, 
and its instantaneous displacement vector, $\Delta \bm{r}(t)$.
Other parameters were set to,  $\hat{\ell_0}=1$, 
$X_\lambda:=D'_{c\beta}/D'_{c\alpha}=1$, 
$X_S=0.01$.}
\label{fig3}
\end{figure}

Fig. \ref{fig3} shows the effect of  $X_r:=\dfrac{k'' }{D'_{\alpha\alpha}\gamma_{\alpha\alpha\beta}}
\sqrt{\dfrac{k_{\rm B}T}{k_b}}$ on the dynamics of the 
Janus dumbbell. As was discussed before 
$X_r$ is the ratio between a characteristic time scale
for the mass transfer of reactants
and a characteristic time scale of the chemical reaction. 
When $X_r>1$ the net reaction rate is limited by the mass transfer of reactant from
the bath to the dumbbell. When $X_r<1$ the net reaction rate is mostly limited 
by the chemical reaction kinetics.
Fig. \ref{fig3}A shows the dimensionless concentrations of the reactant
for three different values of $X_r$ as a function of time. For $X_r=0.01$ the overall reaction 
rate is mostly limited by the chemical kinetics. For $X_r=1$ both mass transfer and reaction
kinetics effects are important. While for $X_r=10$ the overall reaction rate 
is limited mainly by the mass transfer of reactant from the bath to the 
Janus dumbbell. In general, for lower values of $X_r$ the reaction rate is 
slower and therefore it takes longer for the reactant to be fully consumed 
and for the swimming velocity to decay to zero.
Fig. \ref{fig2}C shows the effect of $X_r$ on the mean-squared displacement of 
the center of mass of the swimmer. \textcolor{black}{As pointed out before,
$\langle \Delta \hat{r}_c^2(t) \rangle$ exhibits four characteristic regions.
A diffusive region is observed at short times with $\langle \Delta \hat{r}_c^2(t) \rangle\sim t^{1}$,
this is followed by a ballistic region, where $\langle \Delta \hat{r}_c^2(t) \rangle\sim t^{2}$.
In general, the ballistic region is followed by 
another diffusive region where the diffusivity of the dumbbell is 
larger than the diffusivity in the first diffusive region by a factor of about fifteen. 
At long times, when the reactant 
has been entirely depleted, another diffusive region that is completely
thermal in nature can be observed. This diffusive region at long times exhibits a 
diffusivity that is smaller than the diffusivity observed in the 
intermediate diffusive region by a factor of about 30.} 
For values of $X_r$ smaller than one, the diffusive region that appears at intermediate 
times and in which the dumbbell exhibits the largest diffusivity
is longer than for values of $X_r$ larger than one. This 
is because for $X_r<1$ the net reaction rate is slower and 
the concentration of reactant remains larger 
than zero for a longer amount of time. \textcolor{black}{
In this intermediate diffusive region
there is still a non-zero concentration of 
reactant in the bath but the swimmer can not maintain the ballistic
motion due to tumbling, {\it i.e} reorientation of $\bm{Q}$ due to Brownian forces.
However, as already pointed out the diffusivity of the dumbbell in this region 
can be larger than in the purely thermal diffusive region that appears at longer times 
by a factor of up to thirty.}
Fig. \ref{fig3}D shows the cross-correlation between the 
dumbbell's orientation and its instantaneous 
displacement vector, $S(t)$.
Again $S(t)$ exhibits a maximum at the same times at which 
$\langle \Delta \hat{r}_c^2(t) \rangle$ shows ballistic behavior. 
For all values of $X_r$ the ballistic behavior in the $\langle \Delta \hat{r}_c^2(t) \rangle$
and the maximum in $S(t)$ occur in the same time range, when $\rho_{\alpha, \infty}$
is still very large and the reaction rate is at its maximum .

\begin{figure}[h t]
\center
\includegraphics[width=\linewidth]{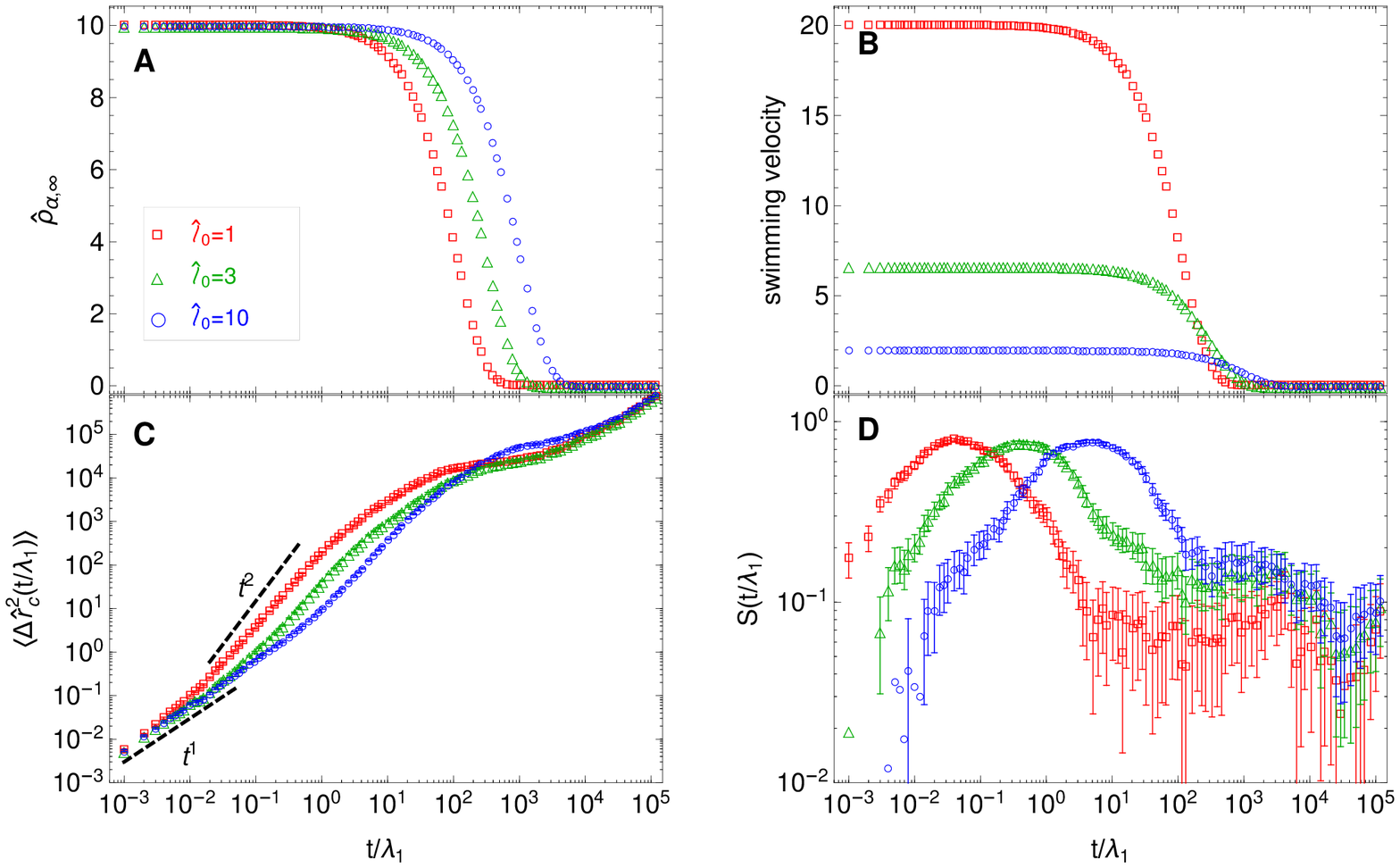}
\caption{Effect of $\hat{\ell}_0$ on the swimming dynamics.
(A) Dimensionless concentrations of the reactant in the bath, 
$\hat{\rho}_{\alpha,\infty}$. 
(B) Dimensionless swimming velocity.
(C) Mean-squared displacement of the center of mass of the Janus
swimmer.  (D) Normalized cross-correlation between the 
orientation of the Janus swimmer, $\bm{Q}/Q$, 
and its instantaneous displacement vector, $\Delta \bm{r}(t)$.
Other model parameters where set to, $X_r=10$,
$X_\lambda:=D'_{c\beta}/D'_{c\alpha}=1$, 
$X_S=0.01$.}
\label{fig4}
\end{figure}

\begin{figure}[h t]
\center
\includegraphics[width=\linewidth]{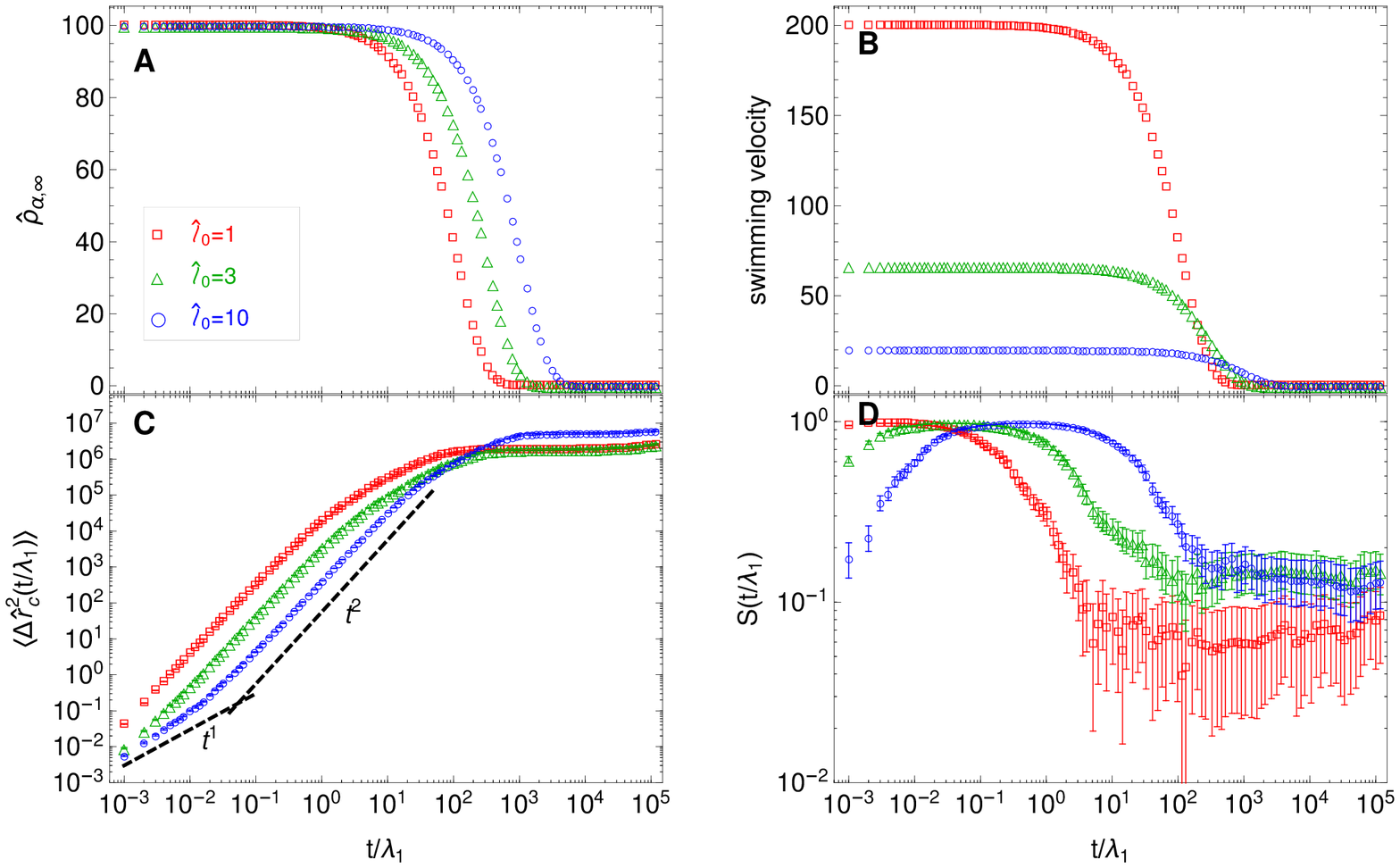}
\caption{Effect of $\hat{\ell}_0$ on the swimming dynamics.
(A) Dimensionless concentrations of the reactant in the bath, 
$\hat{\rho}_{\alpha,\infty}$. 
(B) Dimensionless swimming velocity.
(C) Mean-squared displacement of the center of mass of the Janus
swimmer.  (D) Normalized cross-correlation between the 
orientation of the Janus swimmer, $\bm{Q}/Q$, 
and its instantaneous displacement vector, $\Delta \bm{r}(t)$.
Other model parameters where set to, $X_r=10$,
$X_\lambda:=D'_{c\beta}/D'_{c\alpha}=1$, 
$X_S=0.01$.}
\label{fig5}
\end{figure}

Figs. \ref{fig4} and \ref{fig5} show the effect of the rest length of the
swimmer, $\ell_0$, on the swimming dynamics of the Janus 
dumbbell. Fig. \ref{fig5} corresponds to a larger initial concentration of reactant
and Fig. \ref{fig4} is for a lower initial concentration of reactant.
It can be observed in Figs. \ref{fig4}A and \ref{fig5}A
that for smaller values of $\ell_0$ the reactant is consumed faster. 
That is because the concentration gradients around the swimmer
are larger for shorter Janus dumbbells. These larger concentration
gradients around the shorter 
dumbbells also produce larger swimming velocities as can be observed
in Figs. \ref{fig4}B and \ref{fig5}B. 
This larger swimming velocities for the shorter Janus 
dumbbells are reflected in overall larger 
mean-squared displacements of their centers of mass. \textcolor{black}{
In Fig. \ref{fig4}C the mean-squared displacement of the center of mass 
of the swimmer again exhibits four distinct regions. 
A diffusive region at short times,
followed by a ballistic region. Then a second diffusive region 
in which the swimmer's diffusivity is larger than
in the first diffusive region by a factor of about 15. When the 
concentration of reactant becomes very low, but is still finite,
a plateau is observed. And finally, when the 
reactant is completely depleted, the diffusive region of purely thermal nature
appears. In Fig. \ref{fig5}C it
can be seen that when the initial 
concentration of reactant is made larger by a factor of ten the ballistic region in
$\langle \Delta \hat{r}_c^2(t) \rangle$ broadens,
and for the smaller values of $\ell_0$ the diffusive region at short times 
is not present.}

In Figs. \ref{fig4}D and \ref{fig5}D it can also be observed that 
the region at which $\langle \Delta \hat{r}_c^2(t) \rangle$  exhibits 
the ballistic region coincides with the region at which the 
cross-correlation between the orientation of the dumbbell, $\bm{Q}/Q$, 
and its instantaneous displacement vector exhibits a maximum. When
$\ell_0$ is increased the ballistic region in $\langle \Delta \hat{r}_c^2(t) \rangle$
moves to longer times and so does the maxima in $S(t)$. Note that
for the cases with larger initial concentration of reactant, shown in
Fig. \ref{fig5}D, the maxima in $S(t)$ are much broader, encompassing several
decades, this coincides with also a much broader ballistic 
region in the $\langle \Delta \hat{r}_c^2(t) \rangle$.
\textcolor{black}{Note also that in Fig. \ref{fig4}C
due to the smaller values of the initial reactant bulk concentration, $\rho_{\alpha,\infty}(0)$,
the plateau in the mean-squared displacement 
of the center of mass of the dumbbell is narrower compared to 
whats observed in Fig. \ref{fig5}C. This is because the 
total decoupling of the $\bm{r}_c(t)$ dynamics from the 
$\bm{Q}(t)$ dynamics due to reactant depletion occurs at shorter times 
for lower initial reactant concentrations. 
When $\rho_{\alpha,\infty}(0)$ is made ten times larger as shown in Fig. \ref{fig5} this 
total decoupling takes more time and therefore the plateau in 
$\langle \Delta \hat{r}_c^2(t) \rangle$ is broader and the final 
diffusive region of purely thermal nature appears at longer times, 
that are not shown in the plot.}
\textcolor{black}{Similar behavior in the mean squared displacement 
of the center of mass of diffusiophoretic microswimmers has also been 
observed in more detailed simulations \cite{thakur2011dynamics, LugliJPCC2012}.
Moreover, more detailed simulations 
have also shown that the orientational correlation function attains
 its maximum value when 
the motion of the microswimmer is ballistic \cite{thakur2011dynamics}.
An advantage of the the coarser model introduced in this
work is that it allows for a wider window of time scales 
to be observed at a lower computational cost.}

\begin{figure}[h t]
\center
\includegraphics[width=\linewidth]{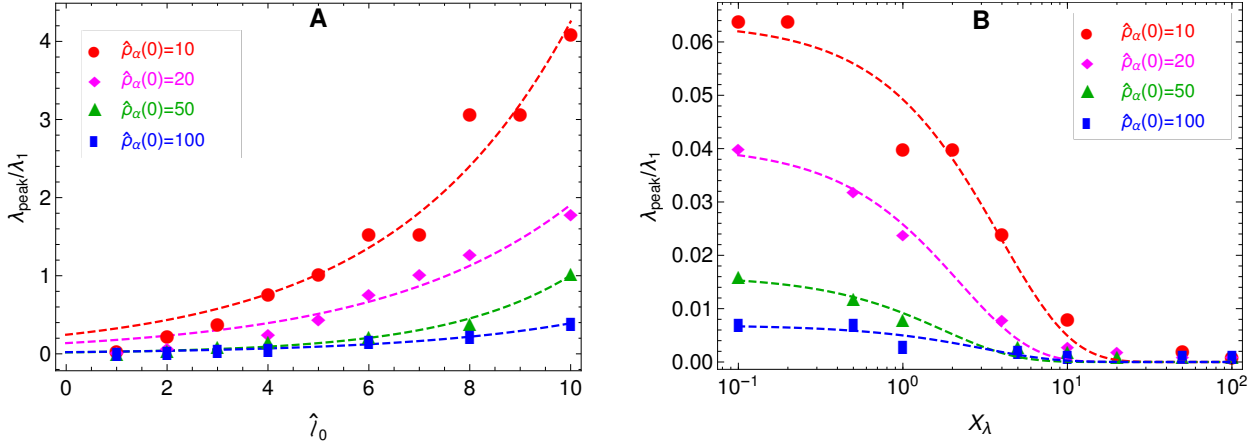}
\caption{\textcolor{black}{Time scales at which the 
cross-correlation between the 
orientation of the Janus swimmer, $\bm{Q}/Q$, 
and its instantaneous displacement vector, $\Delta \bm{r}(t)$,
exhibits a maximum.
(A) As a function of the dumbbell rest length, $\ell_0$, for different values 
of the the initial concentration of reactant in the bulk, 
$\rho_{\alpha,\infty}(0)$.
Other parameters set to
$X_\lambda:=D'_{c\beta}/D'_{c\alpha}=1$, 
$X_r=10$, $X_S=0.01$
(B) As a function of 
$X_\lambda:=D'_{c\beta}/D'_{c\alpha}$
for different values 
of the the initial concentration of reactant in the bulk, 
$\rho_{\alpha,\infty}(0)$.
Other parameters set to $\hat{\ell}_0=1$,  $X_r=10$, $X_S=0.01$.}}
\label{fig6}
\end{figure}

\textcolor{black}{The model presented here may be used as a tool when designing or 
analyzing experiments with Janus swimmers. For instance, Fig. \ref{fig6}
shows the time at which the cross-correlation between the 
orientation of the Janus dumbbell, $\bm{Q}/Q$, and its displacement vector,
$\Delta \bm{r}_c$, exhibits a maximum, $\lambda_{\rm peak}$. 
This is the time at which the Janus dumbbell will 
exhibit swimming dynamics, or in other words the time at which it will 
be able to sustain ballistic motion with a persistent direction. 
The two model parameters, $\ell_0$ and 
$X_\lambda:=D'_{c\beta}/D'_{c\alpha}$, for which the location 
of the maximum in $S(t)$ exhibits dependence have been considered.
Fig. \ref{fig6}A shows how the location of the maximum value of $S(t)$, $\lambda_{\rm peak}$,
depends on the rest length of the dumbbell, $\ell_0$, for different values 
of the initial concentration of reactant in the bulk,
$\rho_{\alpha,\infty}(0)$, for a system
where $D'_{c\alpha}$ and $D'_{c\beta}$ are the same,
\textit{i.e.} $X_\lambda:=D'_{c\beta}/D'_{c\alpha}=1$.
It can be observed that $\lambda_{\rm peak}$ increases exponentially
as $\ell_0$ is increased and decreases exponentially as 
$\rho_{\alpha,\infty}(0)$ is increased. Longer swimmers 
produce smaller reactant gradients across their length and therefore 
are slower achieving swimming motion.
The results in Fig. \ref{fig6}A can be collapsed into a single 
master curve, $\lambda_{\rm peak}/\lambda_1=0.3 
\exp\left({0.3 \hat{\ell}_0}-0.05\hat{\rho}_{\alpha,\infty}(0)\right)$.
In Fig. \ref{fig6}B the dependence of $\lambda_{\rm peak}$
on the ratio, $X_\lambda:=D'_{c\beta}/D'_{c\alpha}$, 
for different values of the the initial concentration of reactant in the bulk, 
for a system with $\hat{\ell}_0=1$ is illustrated. In this case, 
$\lambda_{\rm peak}$ decreases exponentially
as $X_\lambda$ is increased. The swimming motion will
occur at shorter times when $D'_{c\beta}$
is larger than $D'_{c\alpha}$. Or in other words, when 
the diffusivity of the reactant is smaller, the concentration 
gradients across the dumbbell are steeper and the swimming 
motion is observed at shorter times. However
by comparing Fig. \ref{fig6} A with Fig. \ref{fig6} B is evident that 
the effect of $\ell_0$ on $\lambda_{\rm peak}$ is more 
pronounced than the effect of $X_\lambda$. 
The results in Fig. \ref{fig6}B can also be collapsed 
into a single master curve, $\lambda_{\rm peak}/\lambda_1=0.1 
\exp\left({-0.4 X_\lambda}-0.04\hat{\rho}_{\alpha,\infty}(0)\right)$.
These type of master curves could be 
helpful to rapidly determine the time scales at which 
swimming motion should be observable in an experiment  for a given set of values of the initial
reactant concentration, the swimmer's size and/or diffusivity of reactants/products.}

\section{conclusions}

In this manuscript a simple mean-field model for a Janus microswimmer has been proposed. 
The model was inspired by the nonequilibrium thermodynamics 
of multi-component fluids that undergo chemical reactions in the context of the
GENERIC (general equation for the nonequilibrium reversible-irreversible coupling) 
framework. A Janus dumbbell was used to model
the swimmer. The Janus dumbbell was assumed 
to be immersed in a bath of reactant. One of the beads of the dumbbell
catalyzes a chemical reaction that consumes the reactant and produces 
a local gradient across the dumbbell's orientation.
The thermodynamic compliance of the 
swimmer model considered here can be rigorously but easily checked by 
tracking the entropy production rate of the closed system. 
Moreover simple evolution equations for the center of mass and the 
orientation of the Janus dumbbell were obtained.

The mass balance equations for the Janus swimmer were written in the 
phase space as a system of stochastic differential equations
and Brownian dynamics simulations of the swimmer were performed. 
The simulations show that the simple mean-field
swimmer model can exhibit ballistic motion when
the concentration of reactant in the bath is large. 
Moreover, the cross-correlation
between the direction of the swimmer's motion and its orientation
has a maximum at the same times at which the ballistic
motion is observed. The mean-field Janus
swimmer model has a small number of adjustable parameters. 
These are related to the diffusion coefficients of reactant and product, the
rate constant for the surface reaction in the catalytic bead, 
the size of the beads and the rest length of the dumbbell.
Moreover, the entropy balance for the model allows
one to check that the entropy production rate stays positive for a given
set of parameters. In this way it can be guaranteed that 
the the second law of thermodynamics is always 
obeyed in the simulations.
\textcolor{black}{In general, the simulations show that 
in the Janus dumbbell, 
the time scale at which swimming motion is observed
increases exponentially with the dumbbell's length and 
decreases exponentially with the ratio between the diffusivity 
of product and reactant. Moreover,
reactions in which the product has larger diffusion
coefficient than the reactant produce faster swimming.  
Also, shorter Janus swimmers produce steeper gradients 
across their lengths and will therefore swim faster than longer swimmers.
The model presented here can help guide the design and analysis 
of experiments by providing fast estimates of the time scales at which 
swimming motion should be expected to be observed for a given system.}

It is worth noting that the procedure that was followed in this work to obtain a mean-field
Janus swimmer model for which the thermodynamics can systematically
be checked can, in principle, also be extended to systems of several interacting swimmers. 
For instance, instead of assuming that the Janus dumbbell is in a bath of reactants and using
the local gradient approximation used here it is also possible to keep the
full mass balances for the reactants and products. That approach will require the evolution
equations for the swimmer to be solved simultaneously with the full mass balance
equations for the reactants and products.

\section*{Supplementary Material}

\noindent Additional mathematical details
to obtain the equations for the model used here
are given in the supplementary information file.

\section*{Acknowledgments}

\noindent A.C. acknowledges financial support from CONICYT under FONDECYT 
grant 11170056. This work was completed in part with computational resources provided by
the Chilean National Laboratory for High Performance Computing (NLHPC, ECM-02).

\end{document}